# Integrating hidden information which is observed and the observer information regularities

Vladimir S. Lerner, *Marina Del Rey, CA 90292, USA,* lernervs@gmail.com

*Abstract*

The Bayesian integral functional measure of entropy-uncertainty on trajectories of Markov multi-dimensional diffusion process is cutting off by interactive impulses (controls) [1].

The entropy fractions, satisfying minimum of each cutoff maximum, evolve in microprocess, superimposing and entangling the conjugated fractions, which reduces the cutoff entropy and encloses the captured complimentary entropy fractions as a source of an information unit. The impulse step-up action launches the unit formation, while its step-down action finishes it and brings energy from the interactive jump. This finite jump transits from uncertain Yes-logic to certain-information No-logic, transferring the entangled entropy to the forming information unit whose measuring at the end of cutting action kills a final entropy-uncertainty. This control breaks down the entangled correlation that limits information of the unit and memorizes the unit by erasing it information, while spending an amount of energy brought with the control. The Yes-No logic holds Bit-Participator in its forming as elementary information observer analogous to Wheeler Bit created without any physical pre-law.

The control opens access for potential interaction of each discreet information unit with environment and other units, whose entanglement or cooperation yields free information, collected prior to the units' cooperation, which is freeing at decorellation. Binding two units in a doublet and cooperating its opposite directional information unit forms a triplet as a minimal cooperative stable structure.

The information path functional (IPF) integrates the multiple information units of cooperating doublets-triplets, bound by the free information, in gradual information flow of a physical macroprocess.

The macroprocesses, formed according to minimax, sequentially decreases the entropy and maximizes information by enfolding the sequence of enclosing triplet structures of the information network (IN).

The IN couples the doublets in triplets while minimizing their bound free information and rising information forces that enfold the ordered sequence enables attracting new information unit by freeing the information.

While the IPF collects these information units, the IN performs logical computing using the doublet-triplet code. Such operations with the entangled memorized information units, model a quantum computation.

The operations with classical information units, cooperated from quantum information units that IN runs, model a classical computation. The IN different hierarchical levels unite logic of quantum micro- and macro- information processes, composing quantum and/or classical computations.

The observation minimax information law originates and integrates these regularities generating the information observer.

**Introduction**

Human brain integrates information collected from different sources, which are observed, and produces an integrated unique image needed for an observer during interactions with these sources at micro- and macro levels.

Physical approach to the observer, developed in Copenhagen interpretation of quantum mechanics [1-3], requires an act of observation, as a physical carrier of the observer knowledge, but this role is not described in the formalism of quantum mechanics.

A. Weller has included the observer in wave function [4], while according to standard paradigm: Quantum Mechanics is Natural [5-6].

The concept of publications [7,8] states that *in* Quantum Bayesianism, which combines quantum theory with probability theory, 'the wave function does not exists in the world-rather it merely reflects an individual's mental state.'

We have shown (in [9-16]) and in this paper that quantum information processes (rather than Quantum Particles), resulting from Bayesian form of entropy integral measure, arise in *observer* at conversion of the process hidden uncertainty to the equivalent certainty-information path functional measure.

We study the *information* nature of these processes and describe emerging observation and its observer from multiple interactions of random process, which formalizes description of different human communications, physical, biological, cognitive, social, and economical processes *independently* of their origin.

The interactions include both natural Markov process and artificially generated probing-searching actions, for example, in computer-Human communications.

Each interaction *transfers a priori* probabilities to its *a posteriori* probabilities through a stochastic multi-dimensional process whose trajectories alternates this probability's sequence over the process.

The logarithmic ratios of these axiomatic probabilities, integrated along the process trajectories, measures the process' probabilities' distinctions as Bayesian integral entropy-uncertainty of the process interactive connections, which hold this integral measure along the process.

The natural interactions may routinely implement the conversion of some pre-existence uncertainty to post interactive certainty, measured through related pre-interactive (a priori) and post-active (a posteriori) Bayesian probabilities. This concurs with Kolmogorov's law 0-1 of random events' occurrences [17], allowing to represent each interaction by an impulse with No-Yes or Yes-No discrete actions.

Reaching a maximum of the process elevated posteriori probability in the Bayesian integrated uncertainty reveals the conversion of process hidden uncertainty to its certainty.

An interaction, reaching certainty –information in some process' dimension, can be stopped, for example, at an interface with some border-obstacle, which ends (kills) this dimension. Formally, such limitations on Markov diffusion process impose various boundary conditions, which include the process cutoff.

The Bayesian Entropy measure (EF) integrates hidden uncertainties of the process, while each cutoff reveals hidden information which Information Path functional (IPF) measure integrates on the trajectories.

Considering each interaction as interactive action of a potential observer and a sequence of the impulse discrete actions of its observable process, testing and measuring the process uncertainty-Bayesian entropy, allows introduce an observing process as a certain process, converted from the observable process while the information observer appears with releasing information at the conversion of uncertainty to certainty.

In such approach, the impulse Yes –No-Yes-No discrete actions, being a part of random observable uncertain process, are also random and can be fractions of a natural process under observation, while their impulse presents a virtual cutoff associated with the interacting potential observer. The impulse implements the Bayesian probability's inference by cutting virtually, or actually each a posteriori process' fraction.

At the conversion of the hidden entropy to a certain process, the last discreet action of these impulses sequence becomes the certain-real control, which really cuts the uncertainty, releasing information and its observer.

In Markovian model of a physical process, the impulse transforms the Markov process' kernel maximum entropy (of the pre-action) to information (of the post-action), where the entropy is automatically minimized, and the impulse impact, concurrently killing the pre-action, produces the minimum of the maximal post-action, while multiple impulse impacts bring their real information into the IPF.

Such physical process enables creating an objective observer via the actual cutoff actions, which naturally provide the minimax principle of revealing information.

As information grows, crossing some border condition [15], the objective observer becomes a subjective observer, which can intentionally test observable process by interactive artificial impulses and verify collected measurements.

The impulses cut off the minimum of maximal measure convert the observable external process to observer optimal internal information micro- and macrodynamics.

This minimax information law establishes minimum of maximal extraction and minimizes maximal consumption of information as mathematical variation law whose extremal equations determine and predict the process micro- and macrodynamics, information structure of the observer, and functionally unify its regularities.

Each impulse cut initiates elementary (micro) process from two cutting conjugated fractions of the EF uncertainty, which tends to further minimize their cutting minimax by entangling this fraction according to the minimax law.

The entangled entropy (uncertainty) of these dependable (conjugated) processes produces information and its observer after killing the entropy through erasure, whose energy delivers the certain control from that impulse.

This finite control simultaneously breaks down the entangled process' correlation, opening interaction with environment, which allows flow of energy to a forming information unit-primary observer, and limits the unit information while measuring it by cutting the unit fraction of the IPF.

The entanglement might connect two (or three) distinguished fractions with superimposing ebits, whose measurement memorizes their information as elementary doublet (or triplet).

Sequential certain logic connects them in information microprocess.

Potential (uncertain) logic belongs to sequential test-cuts before appearance of initial certain logic, which is a part of forming the elementary information unit with Yes-No certain logic.

Such information unit is a Wheeler Bit whose Yes-No logic forms Bit-Participator, as primary information observer, without any a priori physical law.

Multiple units of the elementary information microprocess, trying to minimize their information, tend to spend their information and energy for cooperating in observer's macroprocesses and macrostructures.

A doublet binds minimal number units by attracting the opposite information units (like that rotating in opposite directions). Free doublet's information-energy, not spent on cooperation, holds a control function, enables attracting other units and structures in building a next minimal cooperative unit-triplet.

The attraction is possible for complementary opposite undistinguished units, having nearest information speeds at forming their cooperation.

These require preparing a ranged manifold of the ordered and complimentary units prior binding them in the cooperation, which enable such manifold the subsequent automatic attraction and cooperation according to minimax.

The rotation mechanism [16] arising at a time-shift of the distributed observable and information processes, orders the distributed units and sequentially cooperates the manifold through the switching control which becomes curved.

The observable process conversion includes multiple probing trials, the measure verification, memorizing information, cooperation in doublet-triplets, and their enfoldment in information network (IN).

Each triplet generates three symbols from three fractions of the information dynamics and one impulse-code from the control. This control joins all three in a single unit and transfers this triple to a next IN triple, forming subsequent hierarchical level of the information network's code-logic. That enfoldment maximizes each cooperating information and concentrates the maximums in the IN nested information geometrical structure, while its free information allocates a feedback path to observation.

In the IN hierarchy, each forming IN high-level logic requests, selects and attracts needed quality of the level information, coordinated through the highest density-frequency of the observer information processing and requesting.

Each information unit has its unique position in the time-spaced information dynamics, which defines the observer scale of both the time-space and exact location of each triple code in the IN and its memorization.

Even though the code impulses are similar for each IN triplet, their time-space locations allow the discrimination of each code with its logic and retraction of this information.

These functional regularities create the united self-operating information mechanism whose integral logic transforms multiple interacting uncertainties to the observer self-forming inner dynamic and geometrical structures within a limited boundary being shaped by the IN information geometry during time-space cooperative processes, and finally originate the intelligent observer's cognitive logic enables the self-programming and predictive computation.

The following sections detail this conceptual description and apply to Brain processing illustrated in the examples.

**1. Observing random uncertain process and its integral measure**

Suppose a manifold of various spontaneous *occurrences* represents a multi-dimensional interactive random process, distributing in space-time, for example, earthquakes, instantaneous changes in stock market, or atomic explosion. Theoretically, a field of multiple distributed probabilities initiates the random process with alternating a priori –a posteriori probabilities along the process trajectories, where each transfer from priori to posteriori probability distribution follows from interaction along the multi-dimensional process (Fig. A). Logarithmic ratio of these probabilities measures relative Shannon's entropy-uncertainty applied to the process' separated states-events: $s_{ab}^s = -\ln[P_a(\delta_i(\tilde{x}_t))/P_p(\delta_k(\tilde{x}_t))]$, where $P_a(\delta_i(\tilde{x}_t)) = P[\delta_{i-1}(\tilde{x}_t)/\delta_i(\tilde{x}_t)]$, $P_b(\delta_k(\tilde{x}_t)) = P[\delta_k(\tilde{x}_t)/\delta_{k-1}(\tilde{x}_t)]$ are a priori and a posteriori probabilities measure at the $i-1, i, k-1, k$ localities $\delta_i(\tilde{x}_t), \delta_k(\tilde{x}_t)$ separating the process events-states. This entropy also measures distinction of the states, bringing to zero each equivalent non-distinguished transformation of the probabilities that does not contribute to the uncertainty measure of randomness.

In a process, moving by chain of events (for example, a metal chain as a sequence of connecting rings), any separation of a state-event-ring dissolves the chain and looses uncertainty transferred by their connection.

Bayesian conditional entropy of the state $s_{ab}^B = -\ln[P_a(\delta_i(\tilde{x}_t))/P_p(\delta_k(\tilde{x}_t))]P_a(\delta_{k+1}(\tilde{x}_t))$ connects this ratio with a probability of the following process' state, which it's priori probability $P_a(\delta_{k+1}(\tilde{x}_t)) = P[\delta_k(\tilde{x}_t)/\delta_{k+1}(\tilde{x}_t)]$ intervening in the following process state's location as a causal probability.

At a very small time-distant of localities $\delta_k(\tilde{x}_t) - \delta_i(\tilde{x}_t) = o(t)$, that a priori probability approaches a posteriori probability of the state $P_p(\delta_k(\tilde{x}_t))$ along the process trajectories.

The entropy measure in the form $s_{ab}^B = -\ln[P_a(\delta_i(\tilde{x}_t))/P_p(\delta_k(\tilde{x}_t))]P_p(\delta_k(\tilde{x}_t))$ allows prognosis the ratio through each transformed posteriori probability $P_p(\delta_k(\tilde{x}_t))$ inferring in the process' connected states using its current a priori-a posteriori probabilities which are alternating along the process.

This leads to process integral measure as Bayesian integral entropy:

$$S_{ap} = -E_{\tilde{x}_t}\int_{s,T}[\ln(P_a(d\tilde{x}_t)/P_p(d\tilde{x}_t))]P_p(d\tilde{x}_t), \qquad (1)$$

where the integral, taken during the process time interval $t \in (s,T)$ and averaged along process $\tilde{x}_t$, describes an integral uncertainty of the process, brought by all hidden randomness through each infinitely small instant as the process' differential $d\tilde{x}_t$.

The process' multiple state connections, changing a nearest probability transformation, automatically integrate the transformation (interaction) of these alternating probabilities along the process trajectory.

Some spread of these interactions (that might differ even in the dimensions) we define as an observable (virtual) process of a potential observer.

In a set of process' events occurrences, each occurrence has probability 1, which excludes others with probability 0. This partitioning is known as the Kolmogorov 0-1 law [17], while their experimental multiple frequencies enable predicts axiomatic Kolmogorov probability, if the experiment manifold satisfies condition of *symmetry* of the equal probable events [18].

Thus, the observable process' multiple interactions in a form of impulses hold virtual probing Yes-No (1-0) actions, whose multiple frequencies enable generating both a priori and a posteriori probabilities, their relational probabilities and the uncertainty integral measure (EF) for each observable process dimension (Fig.B).

Such interactions could be both a natural Markov process and artificially generated during the observable process probs, while the interactive connections hold integral measure of hidden uncertainties (entropy) along the process.

## 2. The impulse max-min cuts off the integral entropy measures and information path functional

Each interaction involves the impulse virtual cutoff action that provides a fraction of uncertainty, which the observable process, with multiple probing impulses, integrates along the manifold of process trajectories as *its* uncertainty-Bayesian entropy's functional measure (EF).

The impulse, cutting process on its separated states, decreases the quantity of process measure (EF) by the amount, which was concealed (hidden) in the connections between the separate states. This measure accumulates more process information than the sum of Shannon's entropies, counted for all process' separated states and confirms a non-additivity of the EF measured process' fractions [12, 16].

The impulse switches the cutting fractions from its EF minimum to maximum and back from the maximum to minimum, holding principle of extracting maxmin-minmax of the EF measure.

Each probing impulse cuts a maximum from each minimal uncertainty, and the control action, revealing its certain *a posteriori* probability, produces maximal information from this minimum, optimizing the extraction of information.

Since the maxmin probes automatically minimize current entropy measure, its actual measurement is not necessary as the minimax cut establishes an information law without any physical pre-law.

The EF variation equations [10,11] predict the optimal cutoff multiple impulses actions, carrying the alternating sequence of a priori- a posteriori probabilities on the extremal trajectories, which implement Bayesian inference actions with these probabilities toward their grow along the process.

As the impulse cutoff enables dissolve the process correlations, it virtually cuts the connection of the process' couple states (at the cutting instance) that frees the maximal entropy of these coupled states, while the cutting fraction holds that maximum from the minimal cut of maximal entropy.

The virtual impulse enable grasp max-min uncertainty through probing frequency measure.

The information, extracted by the numerous control impulses $\delta i_{ap}(d\tau)$ (during cut off interval $d\tau$), integrates it in the information path functional (IPF):

$$I_{ap} = \int_{x_t,T} \delta i_{ap}(d\tau) \qquad (2)$$

along the multi-dimensional observing process $x_t$ during its time $T$.

At a limit, when the cutoff applies to each process correlation, the IPF coincides with EF when observable process $\tilde{x}_t$ approaches observing process $x_t$:

$$\lim_{x_t \to \tilde{x}_t} I_{ap} = S_{ap} \quad . \qquad (3)$$

In information observer, it happens when observable process, being killed by the controls, converts to observation-information process.

Both EF and IPF have exact mathematical descriptions [10-13]-for Markov diffusion process, being a common model of multiple random interactions, as well as the observer process, described by the EF-IPF exremals.

## 3. Information microprocesses initiated by the cutoff

The impulse' step-down action, intervening into observable process, cuts maximal entropy from the process *random* ensembles. The impulse step-up action cuts minimum of this maximum, parting two opposite fractions of the cutting ensemble with maximal distinction of the difference in their max-min entropy measure within this cut (Figs.1,2).

This ensemble fractions' cut carries the related opposite step-up and step-down actions, which start the conjugated dynamic process to minimize a difference of the entropy measure for the conjugated fractions.

These impulses' interfering actions on the Bayesian a priori – a posteriori fractions open covariant conjugated dynamics emerging from the covariance (correlation) of cutting fractions for the considered Markov diffusion.

The observable process, moving in space during the cutting time-shift, originates observer's relative rotating space movement toward non-distinction of the entropy complementary measures of the fractions. The process time course-shift moves the observer Bayesian impulse-control intervention and conversion. This transforms the observable external random process to its internal distributed microdynamics process, starting with rotating conjugated functions of the EF fractions for each cut that primary minimizes its initial uncertainty.

The opposite rotation of the conjugated entropy fractions superimposes their movement in the correlated movement, which further reduces their entropy measure. Such correlated movement entangles the rotating entropy fractions in a microprocess minimizing each cutoff entropy.

Thus, reaching the entanglement brings the observer's cutoff maxim-min entropy with the inverse rotating velocities to a second minimize of the maximal entropy, corresponding symmetry of the probabilities, whose repeating does not bring more entropy and limits the trajectories distinction [16].

Incursion of probing actions near and into the entanglement, prior to verification of the symmetry, does not provide measuring and memorizing entropy, being physically unperformed.

When the various virtual measurements, testing uncertainty by interactive impulses along the observable process, reveal its *certain a posteriori* probability, this inferring probability's test-impulse starts converting uncertainty to certainty-information.

While the impulse step-up action sets a transition *process* from the *observing* uncertainty to information, the impulse's certain step-down control *cuts* a *minimum* from each *maxima* of the observed *information* and initiates internal distributed *information* dynamics process, starting both the *information and its observer* (Figs.1,2).

The information dynamics emerge to minimize the EF-IPF distinctness (3), while the functional extremals solve this variation problem [10,11,13] under the cutoff actions.

The entanglement encloses the captured complimentary (conjugated) entropy fractions providing a source of an information unit. The impulse step-up action launches the unit formation, while its certain control step-down action finishes the unit formation bringing memory and energy from the interactive jump.

This finite jump transits from uncertain Yes-logic to certain information No-logic, transferring the entangled entropy of observation to the forming information unit of elementary (quantum) bit.

Only when the certain step-down control starts cutting the entangled fraction *at* the verification, it exposes the unit of information whose measuring and memorizing at the end of cutting action kills a final uncertainty.

Uncertain step-up logic does *not* require energy [19] like the probes of observable-virtual process, or a Media whose information is not observed yet. This potential (uncertain) logic belongs to sequential test-cuts before appearance of a priori certain logic, which becomes a part of forming the elementary information unit.

The entanglement might connect two (or three) distinguished units with superimposing ebits, whose measurement memorizes their information.

The certain information logic of the forming information unit gets flow of energy from the interactive jump, when the control breaks down the entangled correlation, leading to disentanglement that limits information of the unit and memorizes the unit information by erasing it, while spending an amount of energy brought with the control.

Hence, the control cut of a last posteriori action finishes the unit formation, which carries memory and energy.

The entanglement changes maximal entropy speed to its minimum that imposes pre-conditional changes of the dynamic constraints [14] of distributed process, which allow the following cutoff to convert the minimal uncertainty to information unit with energy.

The Yes-No logic holds Bit-Participator in its forming as elementary information observer analogous to Wheeler Bit [6] created without any physical pre-law.

The control opens access for potential interaction of each discreet information unit with environment (observable process) and other units, whose entanglement or cooperation yields free information, collected prior to the units' cooperation, which is freeing at decorellation.

The decorrelation of the cooperated units differs from the disentanglement at forming the unit by quantity of free information.

The distributed rotation diagonalizes, equalizes, orders the sequentially minimized the eigenvalues of the dynamics' conjugated eigenvectors, and applies this mechanism to the already ordered eigenvectors for their binding in collective structures until their minimax is reached during the rotating movement of the observer information process. The curved control, binding two units in a doublet and cooperating its opposite directional information unit, forms a triplet as a minimal cooperative stable structure. Resonance of the equal frequencies of complementary eigenvalues also performs their cooperation in doublet and triplet.

Triplet structure enables shaping both naturally and artificially.

Physical examples of emerging collective phenomena through active rotation or/and resonance are in [20,21].

The experimental results [21a] concur with the formalism of information microdynamics in which the curved rotated entropy functional fractions, as interfering waves, find pathway through a double as well as triple slits.

**4. Forming information macrodymamic process**

The information path functional (IPF) integrates the multiple information units of cooperating doublets-triplets, bound by the free information, in gradual information flow of a physical macroprocess.

Such a flow starts from statistical process, whose multiple frequencies' test discloses quantum microdynamics, whereas the EF mathematical expectations average the random entropy impulses, and the controls coverts them to microdynamics collected by the IPF, while converting EF to IPF.

The IPF extremal trajectories describe the information macroprocess of the flows cooperating macrounits, which connects the averaged quantum microdynamics on the IPF extremals with classical Physical Statistical Physics and Thermodynamics in Observer.

The selective statistics of minimax probes, which minimize probes number, *specify and distinct* the considered observer macroprocess from that is not selective observed.

This leads to three kinds of certain macroprocesses in observer: One that is formed according to minimax with sequential decreasing entropy and maximizing information enfolded in the sequence of enclosing triplet structures, while minimizing their bound free information.

This is a physical thermodynamic process generating the information network (IN) (Fig.3) with sequential information coupling (doublets and their coupling in triplets) and rising information forces that enfold the ordered sequence and enable attracting new information unit by freeing the information.

The second kind brings an extreme to the path functional but does not support the sequential decrease, minimizing information speed by the end of forming the information unit.

Such units might not be prepared to make optimal triple co-operations and generate the IN structures.

This is a physical macrodynamic process, which we classified as an *objective* or such one that is closed to a border [15] with observer's *subjective* process, enables sequentially cooperate in a process *more* than single triplet.

The third kind macroprocess, transforming energy-information with maximum of entropy production, does not support the observer stability, evolvement and evolution through the information network (IN), while, by holding the elementary triplet, forms the objective information observer.

The IMD Hamiltonian linear and nonlinear equations describe the irreversible thermodynamics of all three kinds [10], which include the information forms of thermodynamics flows defined via the IPF gradients as the process' information forces.

Regularities of these processes arise after imposing dynamic constraints [13], which block the initial randomness.

The IN optimal co-operations hold observer's optimal macrostructures, but they are not necessary for a non-optimal sequence of units that forms a physical macrodynamic process, where each next unit randomly transfers energy-information to the following in various physical movements, including Markovian diffusion.

These natural interactions routinely implement the conversion of pre-existence uncertainty to post interactive certainty through pre-interactive (a priori) and post-active (a posteriori) Bayesian probabilities [22,23].

Even though such multiple cooperation could bring an extreme of the path functional, each following unit might not decrease the extremum, while each IN's triplet minimizes the bound information-energy that they enclosed, including triple quantum information.

Implementation of the variation principle requires compensation entropy production for the irreversibility by equivalent negentropy production, synchronized by Maxwell's Demon [24].

On a primary quantum information level, it accomplishes the sequence of Yes-No, No-Yes actions following from the minimax at the cutoff [16]. A priori step-up Yes-control transfers total cutoff entropy (including the impulse's entropy) to equivalent information, and the posteriori step-down No-control kills the entropy while compensating it by equal information of the control. Such control models an interaction coming from natural physical process like earthquake others, as observable process, being uncertain for potential observer.

These natural interactions are a source, creating objective observer with the elementary doublets and triplets.

The triplet, carrying naturally born free information, enables use it for cooperation with other naturally born doublets –triplets. If such cooperation brings information, allowing to overcome the threshold with subjective observer, it starts the subjective observer's IN with ability of free information to attract new triplets' information.

The end of Bayesian a priori interactive action kills the impulse entropy, while its time transition to a Bayesian posteriori inferring action delivers this entropy-information cost for converting that entropy to information bit, which is estimated by coefficient $k_e = s_{ev} / \ln 2 \cong 0.0917554$ [16]**.**

Here, the real posteriori part of the impulse brings information equivalent, which covers entropy cost $s_{ev}$ and brings the real step-up control of a bordered next impulse. If an external step-down control, which is not a part of the impulse, spends the same entropy, and kills a cutting part of the impulse entropy $u_i$, then $k_{eu} = s_{ev} / u_i \approx 0.1272$.

Time interval of the conversion gap $\delta_o$ is $\delta_t^o = u_{oi}/c_{oi}$ where a speed of killing this entropy is $c_{oi}$.

Since by the end of $\delta_o$ the information unit appears with amount $\mathbf{a}_{io} = u_{oi}$, the real posteriori speed $c_{oi}$ produces the finite posteriori control speed. Thus, this is speed of generating information, whose maximum estimates constant $c_{mi} = \hat{h}^{-1} \cong (0.536 \times 10^{-15})^{-1} Nat/\sec$ [13], where $\hat{h}$ is an information analog of Plank constant at maximal frequency of energy spectrum of information wave in its absolute temperature. This allows us to estimate minimal time interval $\delta_{t\min}^o \cong \mathbf{a}_{io}\hat{h} \approx 0.391143 \times 10^{-15} \sec$, which determines the ending border of generation information unit $\mathbf{a}_{io}$. This unit contains an information equivalent $s_{ev} = i_{ev}$ of energy $e_{ev}$ spent on converting the entropy to information. Energy $e_{ev}$ delivers the certain step-up control action of converting entropy to information. This energy compensates for that conserved in the entangled rotation, specifically by angular moment multiplied on angular speed during the time movement [9], which evaluates entropy $s_{evo}$ in Nats. Above ratio $k_e$, measured by the equivalent cost of energy, is a quantum process' analogy of Boltzmann constant (as a ratio of the unit generating radiation energy, which is transferred to temperature of heat that dissipates this energy). The ratio determines a part of total information $\mathbf{a}_{oi}$ cost for transforming entropy $u_{oi}$. This information cost is a logical equivalent of Maxwell Demon's energy spent on this conversion, as the cost of logic, which is necessary to overcome $\delta_o$, including the transferring of the entangled entropy and generation of the unit of information during time interval $\delta_t^o$.

Since the rotating movement condenses its entropy in the volume of entanglement, this logical cost is a primary result of time shift $\delta_t$ initiating the rotation, while this time course compensates for both logical and energy costs, delivering the certain step-wise control. This means, that real time course might be enough for excluding any other costs for conversion entropy in equivalent information.

Moreover, since this time course conserves the entropy of entanglement, the conversion does not require any other real cost at the real time course, which provides the related space. However, initially the time–space is result of observable interactive impulse (Figs.A,B), which is an observer- process' participator (inter-actor) of converting to certainty- information. If it is true, then such interaction is responsible not only for time-space creation, but for non-cost transformation of entropy to equal information-certainty in dissipating observer's current time course.

Information mass [14] of rotating cooperating units, acquiring energy after entanglement, models elementary particles, as well as their various collective formations at the curving knots as the IN nodes (Fig.3),[25].

The step-up cut jumps the curvature of the space-time distribution, initiating an attractive wave of cooperation.

A conjugated quantum process is reversible until it reaches entanglement at superposition (interaction) of the process' complimentary entropy fractions, directed toward the generation of information.

The potential equalization uncertainty-certainty requires the existence of the *extreme* process, heading for both equalization and creation of the information unit with maximum information speed. The extreme condition [9],[11] compensates for entropy production through the VP Hamiltonian, measuring integral speed of the cooperating information process, which actually performs Maxwell's Demon function at macrolevel. The condition of the *minimization* of this speed (imposed by the VP dynamic constraint [13]) limits information-energy for emerging unit by cutting it at formation and freeing the conserved information [16]. Possible continuation of this extreme process in a double pair cooperation, with maximum information speed (entropy production) leads to self-destruction of any created information. If the extreme process' chain of cooperating states satisfies the minimax, the chain's free information spends minimal entropy on sequential pair-cooperation of the states with the minimal information speeds. The chain's generated physical thermodynamic irreversible process encloses information, which connects its states-units through broken symmetry at the information microlevel .

The units of information had extracted from the observed random process through its minimax cutoff of hidden information which observer physical thermodynamic process integrates via Eigen Functional [26,10], satisfying the VP on a macrolevel in the IPF form. The IPF holds the generated primary free information that sequentially connects the pairs of states-units as doublets and the multiple doublets-triplets co-operations.

**5. Arising Observer Logical Structure**

The energy-information units of a primary triplet start an elementary physical information process and compose a minimal logical code that encodes the process. Each triplet generates three symbols from three segments of information dynamics and one impulse-code from the control. This control joins all three in a single unit and transfers this triple to next triple, forming next level of the information network's (IN) code.

The triplet's dynamic connection holds an elementary thermodynamic process with minimum of three such logic structures during the IN formation.

Each information unit has its unique position in the time-spaced information dynamics, which defines the scale of both time-space and the exact location of each triple code in the IN. Even though the code impulses are similar for each triplet, their time-space locations allows the discrimination of each code and its forming logics.

The observer checks the acceptance of this code for its information network (IN)(Fig.3). This includes enclosing the concurrent information in a temporary build IN's high-level logic that requests new information for the running observer's IN [15, *v2*]. If the code sequence satisfies the observer's IN code, the IN decreases its free information by enfolding new code in its structure. The IN information cooperative force, requests this compensating information [12]. The decrease is an *indication* that the requested information, needed for growing and extension of the IN logic has been received, which is associated with surprises that observer obtains from the requested information it needs.

The IN connections integrate each spatial position of the accepted triple code into the triplets' information of the IN node's previous position while structuring the IN. Timing of observer's internal spatial dynamics determines the scale of currently observed process and serves for both encoding and decoding the IN logic.

The IN parameters are identifies by observer's measured frequencies of the delivered information [15].

The spatial coordinate system rotates the following next accumulation and ordering of the IN node's information. Therefore, the dynamics generate the code's positions that define the logics, while the observer information creates both the dynamics and code.

The space-time's position, following from the IN's cooperative capability, supports the *self-forming* of observer's information structure (Fig.3-4), whose self-information has the distinctive *quality measure*.

The observer identifies order and measure of his priorities (like social, economic, others categories) using quality of information concentrated within the IN node, assigned to this priority.

The receiving information of this priority node's quality is directed to other (local) IN, which emanates from that node. This particular information is collected and enfolded according to the sub-priority of competing variation's quantities of each category, by analogy with main IN (Fig.3).

Each category of priority information builds its local IN emanating for the main IN.

When the node quality is compensating by external information, the observer achieves this priority as an unpredicted surprise; if it is not, the observer local IN requests for the new one.

Fig.5. illustrates an example of forming a cooperative triple for local IN that is attaching at locality of the main node.

Timing of current observation depends on the IN logic accumulated in previous time, which demands this observation.

The observer logic with its timing hierarchy might initiate external interaction, following observation, entanglement, generation information units, and the IN nodes, forming its multiple finite sizes knots (Fig.3).

This time interval could be minimizes for the sequential interactions approaching speed of light.

In particular, in human observer's vision, the ability to see and form mental pictures has a speed much higher than external interactions [27].

Finally, the irreversible time course generates information with both objective and subjective observers, which can overcome the information threshold between them on the path to intelligence.

Observer computes its encoding information units in the information network (IN) code-logic to perform its integrating task through the self-generating program. The observing process, which is chosen by the observer's (0-1) probes and the logic integration, could determine such a program. As a result, the information path functional (IPF) collects these information units, while the IN performs logical computing operations using the doublet-triplet code.

Such operations, performed with the entangled memorized information units, model a quantum computation.

The operations with classical information units, which observer cooperates from quantum information units and runs in the IN with these units, model a classical computation. An observer that unites logic of quantum micro- and macro- information processes enables composing quantum and/or classical computation on different IN levels.

That includes integral measuring each observing process under multiple trial actions, converting the observed uncertainty to information-certainty by generation of internal information micro and macrodynamics and verification of trial information. Enclosing the internal dynamics in information network (IN); building the IN's dynamic hierarchical logic, which integrates the observer requested information in the IN code; and enfolding the concurrent information in a temporary build IN's high-level logic that requests new information for the running observer's IN.

# 6. Observer Intelligence

The observer's current information cooperative force [15], initiated by its IN free information, determines the observer's *selective* actions and *multiple choices,* needed to implement the minimax self-directed strategy.

Notion of *conscience* [12,28,29,16] we associate with *selecting* the *optimal* observer's choices through the cooperative force emanated from the integrated node.

The selective actions are implemented via applying the considered stepwise controls with a feedback from the currently checked information density, requesting such spectrum information speed, which conveys the potential cooperation, needed to verify the chosen selection.

The cooperative force includes dynamic efforts implementing the observer's *intention* for attracting new high-quality information, which is satisfied only *if* such quality could be delivered by the frequency of the related observations through the selective mechanism.

These actions engage acceleration of the observer's information processing, coordinated with the highest density-frequency of observing information, quick memorizing each node information, and stepping up the generation of the encoded information with its logic and space-time structure, which minimizes the spending information.

The needed selection is predicted through the time interval [15] of the control action, which is required to select the information that carries the requested information density.

These processes, integrating by the IPF, express the observer cognition, which starts with creation of elementary information unit with memories at microlevel.

The coordinated selection, involving verification, synchronization, and concentration of the observed information, necessary to build its logical structure of growing maximum of accumulated information, unites the observer's organized *intelligence action*, which evaluates the amount of quality of information spent on this action *integrated in the IN node*.

The functional organization of the intelligent actions, which integrates cognition and conscience of the interacting observers, connects their levels of knowledge, evaluated by the hierarchical level's quality information [15] that the observers have accumulated and stored (memorized) in each IN.

Result [30] confirm that cognition arises at quantum level as "a kind of entanglement in time"…"in process of measurement", where… "cognitive variables are represented in such a way that they don't really have values (only potentialities) until you measure them and memorize", even "without the need to invoke neurophysiologic variables", while "perfect knowledge of a cognitive variable at one point in time requires there to be some uncertainty about it at other times".

# 7. The experimental illustrations and confirmation of the observer information regularities on examples from neuronal dynamics. Comments

Analyzing these example, presented in reviews and studies [31-55] (which include extensive References), we explain and comment some of their *information functions* using formal results [9-16] and this paper.

Here, we relate observer's external processes to sensory information, while its internal processes uphold the reflected information internal dynamics' (IMD) synchronized proceedings.

According to review [31], a single spike represents an elementary quantum information, composed from the interfering wave neuronal membrane potentials, which is transferred to post-synaptic potential.

The synaptic transmission is a mechanism of conversion of the spike, or a sequence of spikes, into 'a graded variation of the postsynaptic membrane potential' that encodes this information… The pre-synaptic impulse triggers connection with a synaptic space, which starts the Brownian diffusion of the neuro-transmitter, opening the passage of information between the pre- and the postsynaptic neuron.

The transferring 'quantum' packet determines the amplitude and rate of the post-synaptic response and the type and quality of the conveyed information.

*Instead of the authors' term 'quantum', we use information-conjugated dynamics that convey the same information.*

'Although there are several possibilities to start releasing the states on the border the post synaptic transmission, usually only a single one close to the centre of the Post Synaptic Density have a higher probability to *starts* the opening formation of binding the receptors'. This 'eccentricity' influences the rate of neuro-transmitters that bind the postsynaptic receptors [31].

*It illustrates selectiveness of the specific minimax states of the Brownian ensemble (located on a cutoff border the Brownian diffusion, Sec. 3), allowing to chose the maximum probable states of the ensemble, having the rate, amplitude, and the time course of the post-synaptic potential, which is triggered by the impulse (Fig.1), related to the spike.*

Results [32] proposes probabilistic spiking neuron model and suggests ways of building it with associative memory and quantum algorithm.

According to [33] "Neurons are the basic information processing structures in the central nervous system, nothing *below* the level of neurons does."

*In Sec.3, the elementary information unit with memory arises in process of measurement the probabilistic certain impulse.*

Each Post Synaptic Density is composed of four chains arranged by any combination of four subunits, which have two types of binding receptors [31].

*This is a straight analogue of our four units in each elementary information level of information network (IN)(Sec.4) with two biding actions of the controls (Fig.3), composing the triplet's node, which holds the controls with actual impulse's high and length, and the IN memory.*

The experiment [34] provides "evidence for the analog magnitude code of the triple-code model activating the bilateral inferior parietal lobules, and that the task of magnitude comparison is conducted in this area, not only for Arabic digits but dot representations as well" for "semantic knowledge for numerical quantities on a mental number line".

Investigated in [35] the triple-code model of numerical processing, involving analog magnitude, auditory verbal, and visual Arabic codes of representation, indicates that "the triple code may be the neuropsychological bases of the performance of complex mathematical tasks".

The starting composition of the receptors subunits depends on the level of 'maturation' of the synapse [36], which is changed during the development and as a function of the synaptic activity.

'The real interaction between excitatory and inhibitory synapses is based on delicate equilibriums and precise mechanisms that regulate the information flow in the dendrite tree.

There is a sort of equilibrium between the cooperative effect due to the summation of the synaptic activities of excitatory neurons and the inhibitory effect due to the reduction of the driving force that produces the information transmission'.

*The minimax mechanism includes regulation, equilibrium and cooperation, whose joint proceeding constrains both internal and external information processing, reducing it for the cooperative connections, Sec.4.*

'The post-synaptic regulation of the information, passing by synaptic transmission, is governed by several different mechanisms involving: the diffusion of the neuro-transmitter in the synaptic space, the absolute and relative number of receptors' Post Synaptic Density, their dynamics and/or composition. These functions' depend on geometry of dendric tree, which has hierarchical structure' *analogous to the IN,* (Fig.3). 'Distinctive levels of potentials between different dendrite tree branches produce information flows between the grafting (joint) points of the dendrite tree's branches. Dendrites of the level $A_n$ are embedded onto 'mother branches' that are at the level $A_{n-1}$ that, in turn, are grafted to their mother branches at the level $A_{n-2}$ and so on up to the main dendrite ($A_0$) emerging directly from the starting information. As the synapses are the points of passage of the information among the neurons, the grafting points can be considered the points of passage of information among the dendrite branches. A good approximation to this situation could be a model where each daughter branch is considered a sort of a bidirectional "electrical synapse" of the mother branch: 'the activity of each branch of the level $A_n$ can influence the activity of the branch of the level $A_{n-1}$ and vice versa... The direction of the current depends on the values of the potential level of two sides of grafting point $VA_n$ and $VA_{n-1}$, being in the direction of the mother for $VA_n > VA_{n-1}$, and in the direction of the daughter in the opposite case.' 'The grafting points represent "nodes" for information flowing in the dendrite tree' [31]. This network holds the structural functions of dendrite tree, its hierarchically dependent nodes, bidirectional flows of information, concurrent formation of all structure with starting flow of information, selective maxmin influence of nearest nodes and their decrease with the distance from the node.

*The described structure has a direct analogy of the IN nested structure, holding two opposite directional information flows, which models not only observer's logical structure, but rather real dendrite bran structure during each of its formation.*

'Although it is generally accepted that the spike sequence is the way the information is coded by, we admit that neuron does not use a *single neural code* but *at least two*' [31].

*Here, the authors suppose that there is a spike code (that might hold the IMD controls with actual impulse's high and length), another is the triplet's code, composed of four subunits, which is transferred between IN nodes [12].*

*The control impulse rises at the end of the extremal segment, after information, carried up by the propagating dynamics, compensates the segment's inner information (determined by the segment macroequations, dynamic constraint and invariants). First, this establishes the direct connection between the information analogies of both the spike dynamics and the threshold. Second, it brings the spike information measure for each its generation, evaluated in bits. The interspike intervals carry the encoded information, the same way that the discreet intervals between the applied impulse controls do it. Conductivity of an axon depends mainly on inter-neuronal electrical conductance, which, in IMD model, is determined by the diffusion conductivity of the cooperative connections, computed via a derivation of the correlation functions. A signal, passing through this conductivity, might modify a topology of a single, as well as a multiple connection, changing its macrodynamic functions (and a possibly leading to distinct networks) under different inputs.*

Neural oscillatory networks, measured by the brain oscillation's frequency, power and phase, dynamically reduce the high-dimensional information into a low-dimensional code, which encodes the cognitive processes and dynamic routing of information [37].

*The IMD model is characterized by the sequential growth of the information effectiveness of the impulse and step controls along the IN spatial–temporal hierarchy. This connects with changing the quality of the IN node's information depending on the node's location within the IN geometry. The changes increase the intensity of the cooperative coupling and its competitive abilities, which make the segment's synchronization more stable and robust against noise, adding an error correction capability for decoding [14-15].*

Even though these and other related results were published, the reviewers [31] conclude:

'None of the approaches furnishes a precise indication of the meaning of the single spike or of the spike sequences or of the spike occurrence, in terms of "information" in the symbolic language of the neurons'.

Now is widely accepted [37,38] that brain system should integrate the observed information, using 'working memory'- long-term potentiating (LTD) for a predictive learning.

There are different LTD experimental and simulation studies [38-42] for dynamic integration [43] of observation and verification of dendrite collected information. It's specifically shown [38,39] that a single neuron could be decomposed into a multi-layer neural network, which is able to perform all sorts of nonlinear computations.

The bimodal dendritic integration code [36] allows a single cell to perform two different state-dependent computations: strength encoding and detection of sharp waves.

Key components of integral information processing are the neuron is synaptic connections, which involve the LTD slow-motion information frequency.

"Observed, executed, and imagined action representations can be decoded from ventral and dorsal areas"[44].

*The information observer concurrently verifies the Bayesian entropy integral, until all its portions will be converted in the total integral information(Sec.2).Information, memorized in observer's existing IN level, has less information frequency, compared to the current collection of the requested high-level information (Sec.6). When information is temporary assembled in the local IN (with STM), its reverse action works as an information feedback from the existing to a running level of the IN. If the memory retrieval coincides with requested information, such IN reactive action could enforce extraction of information and its synchronization (working as a modulation in the LTD) with that needed by the observer's IN. That helps integration of both EF and IPF by binding their portions according to the time course, which supports the LTD functions (and a possibly leading to distinct networks) under different inputs.*

The experiments and simulation [42] show that maximizing minimum of the EF integral Bayesian probability information measure (corresponding the authors multiple probabilities) allows to reach the effective LTD learning [41].

The review results [36-40] confirm that the dendrite branches pyramidal neurons function, working as single integrative compartments, are able to combine and multiplicate incoming signals according to a threshold nonlinearity, in a similar way to a typical point neuron with modulation of neuronal output.

According to studies [45,46], an observer automatically implements *logarithmic* relationship between stimulus and perceptions, which Weber's law establishes. Additionally, Fechner's law states that subjective sensation is proportional to the logarithm of the stimulus intensity.

*Both results imply that observer's neuronal dynamics provide perceptibly of both entropy measure and acceptance of information(as a subjective sensation).*

Review [47] provides the evidence of converging the motivational effect on cognitive and sensory processes, connecting attention and motivation.

Results [48] support autonomous cell-cell communication of DNA encoding distinct logic-gate states. And [49] details cell-to-cell communication which coordinates the behavior of individual cells to establish organ patterning and development.

The dynamic brain networks cognitive activities [50-51] demanding attention, as is the case for aesthetic appreciation [49] show the functional connectivity dynamics along three temporal windows and two conditions, beautiful and not beautiful stimuli. Where the experiments support for the aesthetic appreciation relies on the activation of two different networks, an initial aesthetic network and a delayed aesthetic network, engaged within distinct time frames. Activation of the networks regions corresponds mainly to the delayed aesthetic network in the mind/brain interaction.

*These confirm theoretical existence of primary temporary building the local information networks (IN) based on a node of existing IN, and surprise (appreciation) involve activation of both local IN and the delayed main IN along with others INs activated for specific tasks[15].*

*The IMD related sensory-cognition processes build and manage a temporal-space pathway to information stored in the IN [16].*

The reviews [41,52] describe classifying networks based on abstract architectures and functions, rather than on the specific molecular components of the networks to design table of core molecular algorithms (information code) that could serve as a guide for building synthetic networks.

*That confirms independence of information network structure on specific biology of building elements.*

Brain–machine interfaces [51] driven by selected attention associates with cognitive activity, which leads to understanding the cognitive function of interest, spatial attention, working memory, other cognitive signals.

Results [52] demonstrate that different components of intelligence have their analogs in distinct brain networks' functional organization of the brain, based on the higher-order factor, accounted for cognitive tasks co-recruiting multiple networks, which confirm the independence of these components of the individual differences.

Review [53 ] considers `intelligence as goal directed systems' 'through the ability to learn to achieve the goal, associated with cognitive optimization and prediction, suggesting connections to the realms of human subjective experience'.

The multiple interactive actions emerge as a source of these activities.

Brain–computer interfaces [54] involve *multitasking* information and cognitive processes, such as attention, and conflict monitoring.

The largest computer simulation [55] embraces only a small part of real Human Brain processing.

*The cited in [12,15,16]) and many other neurodynamics studies also support and confirm the paper formal results.*

# Figures

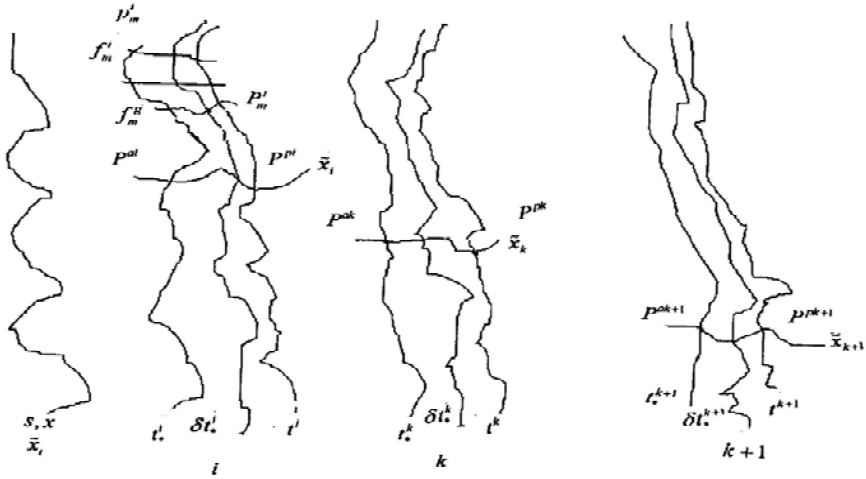

**Fig. A.** Schematic illustration of probability distributions $(i, k, k+1)$ with alternating a priori $P^{ai}$ -a posteriori $P^{pi}$ probabilities, their connection to experimental frequencies $(f_m^i, ... f_m^g)$ and to experimental a priori $p_m^i$ and a posteriori $P_m^i$ probabilities; $\tilde{x}_t(\tilde{x}_i, \tilde{x}_k, \tilde{x}_{k+1})$ is multi-dimensional random process with components $(\tilde{x}_i, \tilde{x}_k, \tilde{x}_{k+1})$ and initial conditions $(s, x)$.

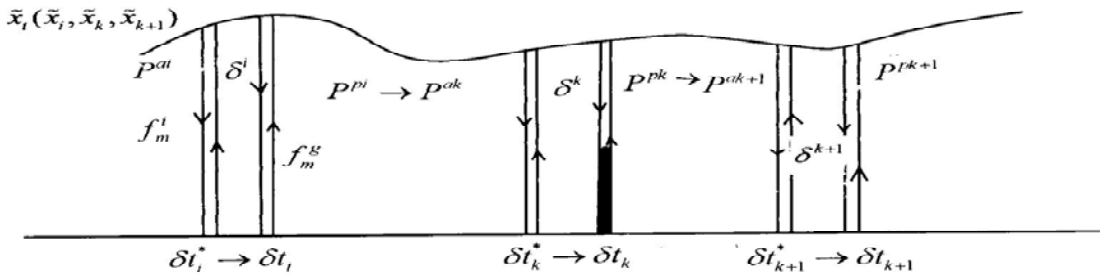

**Fig. B.** Schematic illustration of interactive impulses arising in observable (virtual) process $\tilde{x}_t$ as observer's series of probing action: $\delta^i, \delta^k, \delta^{k+1}$ whose frequencies reveal observer's a priori –a posteriori probabilities $P^{ai} \to P^{pi} \to P^{ak} \to P^{pk} \to P^{ak+1} \to P^{pk+1}$ during time intervals $\delta t_i^* \to \delta t_i, \delta t_k^* \to \delta t_k, \delta t_{k+1}^* \to \delta t_{k+1}$, where each symbol $\to$ indicates the transfer from observable (virtual) time to the observing (certain) time intervals during the probing impulses; within the impulse $\delta^k$ (for the observable process' dimension $\tilde{x}_k$) starts a certain step-up control by index $\uparrow$ (Yes) (shown in bold) with uncertain gap-delay $\delta_o$ (Sec.8).

**Fig.C.** Applying impulse controls (IC), composed of the step-down (SP1) and step-up (SP2) functions.

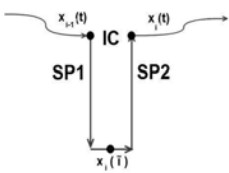

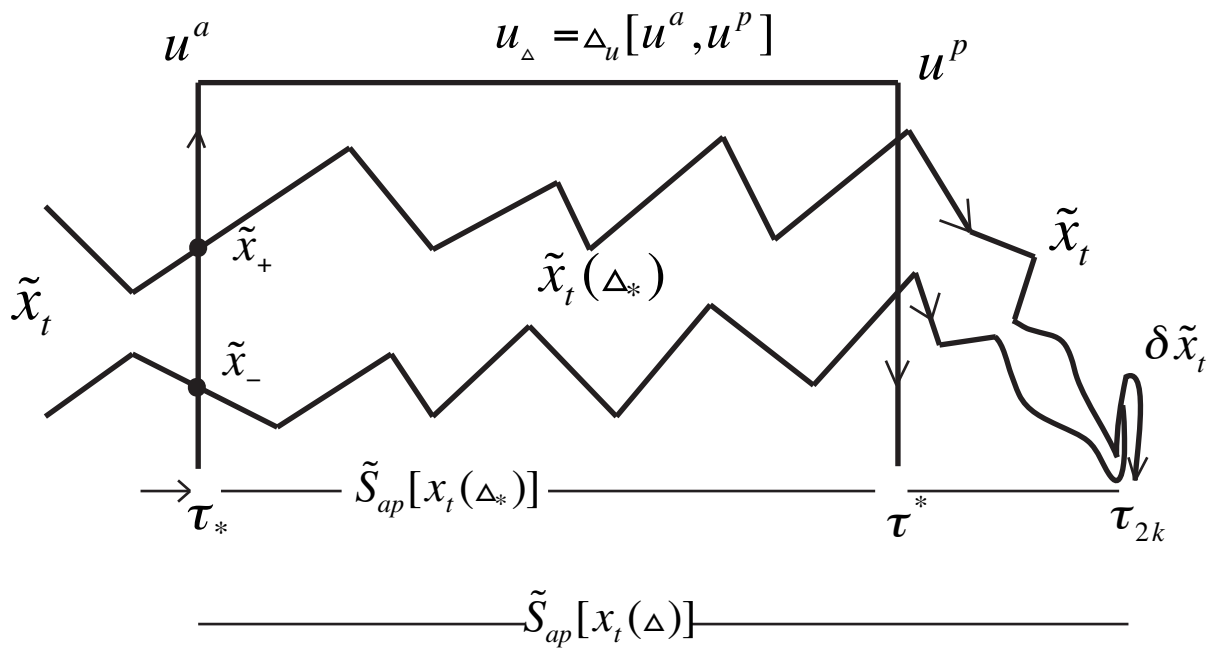
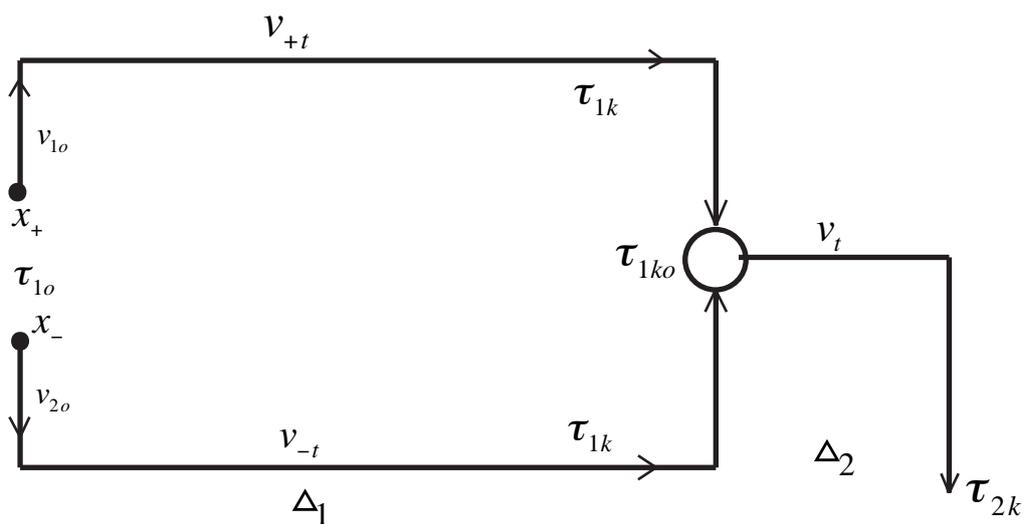
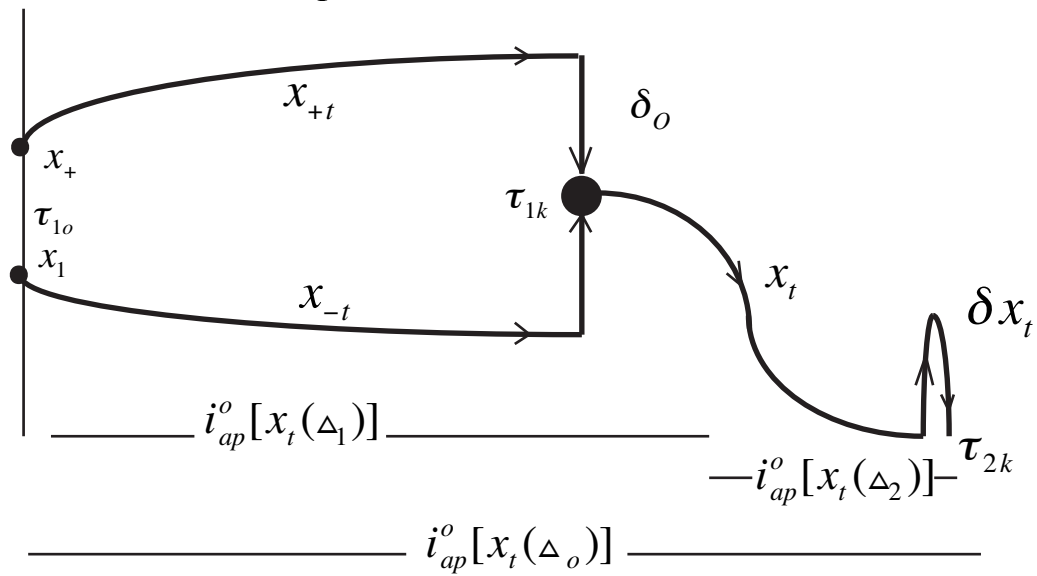

**Fig.1. Illustration of the observer's *simultaneous* proceeding of its external and internal processes and holding information.**

In this Figure: $\tilde{x}_t$ is external multiple random process, $\tilde{x}_t(\Delta)$ is potential observation on interval $\Delta$, which randomly divides $\tilde{x}_t$ on a priory $\tilde{x}_a(t)$ and a posteriori $\tilde{x}_p(t)$ parts; $u_\Delta = \Delta_u[u^a, u^p]$ are impulse control of parts $\tilde{x}_a(t), \tilde{x}_p(t)$; $\tilde{s}_{ap}[\tilde{x}_t(\Delta)]$ is observer's portions of the entropy functional; $\tilde{x}_t(\Delta_*), \Delta_*, u^a(\tau_*), \tilde{u}^p(\tau^*)$ and $\tilde{s}_{ap}[\tilde{x}_t(\Delta_*)]$ are related indications for each cutting process; $x_t(\Delta_o)$ is observer's internal (conversion) process with its portion of information functional $i_{ap}^o[x_t(\Delta_o)]$; $\tau_{2k}$ is ending locality of $\tilde{x}_t$ with its sharp increase $\delta \tilde{x}_t$; $\tilde{x}_-, \tilde{x}_+$ are the cutting maximum information states; $v_o(v_{1o}, v_{2o})$ are observer's opposite inner controls starting with $x_-(\tau_{1o}), x_+(\tau_{1o})$ complex conjugated trajectories $x_{-t}, x_{+t}$ interfering nearby moment $\tau_{1k}$; $v_{+t} = f(x_+, x_{+t}), v_{-t} = f(x_-, x_{-t})$ are inner control functions; interfering nearby moment $\tau_{1k}$; $\delta_o$ is interval of the control switch from $\tau_{1k}$ to $\tau_{1ko}$, where unified mirror control $v_t$ entangles the dynamics on interval $\Delta_2$ up to $\tau_{2k}$- locality of turning the constraint off with sudden rise $\delta x_t$.

The shown external and internal intervals could have different time scale.

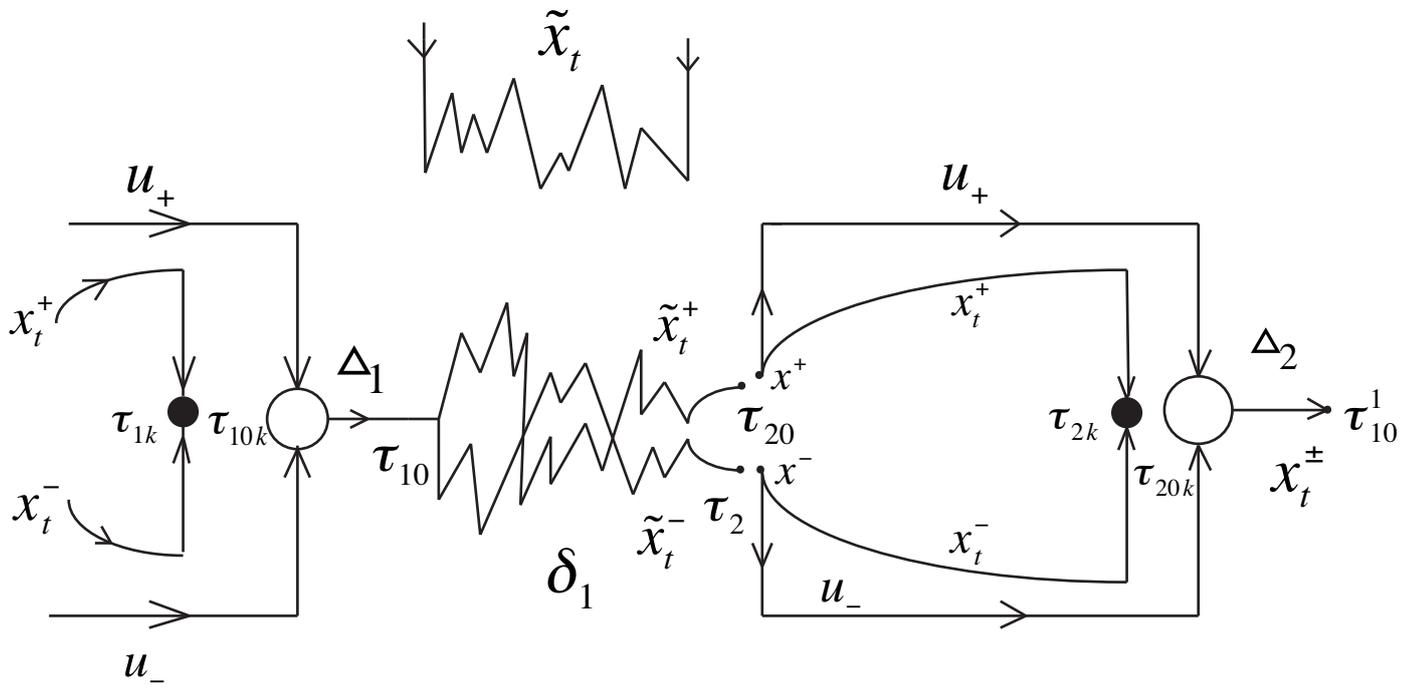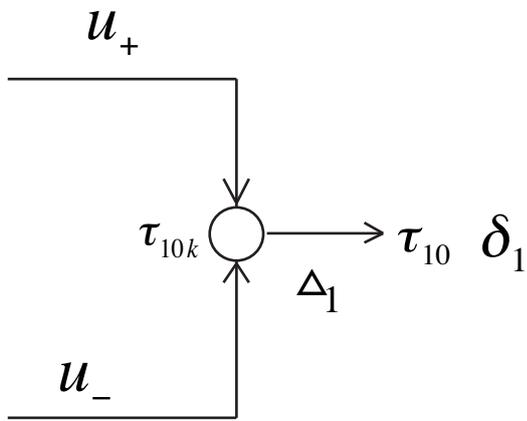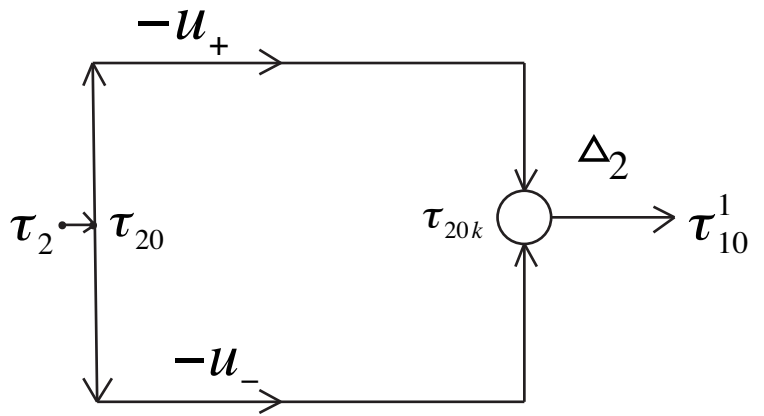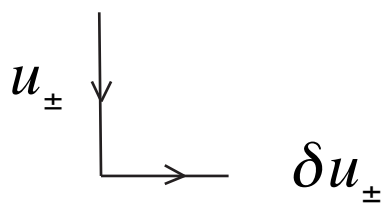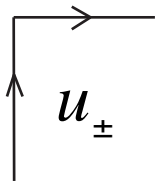

**Fig.2. Illustration of the observer's *sequential* proceeding of its external and internal processes.**

In this Figure: $\tilde{x}_t$ is external multiple random process; $\tilde{x}_t^+, \tilde{x}_t^-$ are copies of random process' components, selected via intervention of the double controls $u_+, u_-$ at the moment $\tau_2$ ; $x_t^+, x_t^-$ are conjugated dynamic processes, starting at the moment $\tau_{20}$ and adjoining at the moment $\tau_{2k}$ ; $\tau_{20k}$ is a moment of turning controls off ; $x_t^\pm$ is adjoint process, entangled during interval $\Delta_2$ up to a moment $\tau_{10}^1$ of breaking off the entanglement; $\tau_{1k}, \tau_{10k}, \Delta_1, \tau_{10}$ are the related moments of adjoining the conjugated dynamics, turning off the controls, duration of entanglement, and breaking its off accordingly, -in the preceding internal dynamics; $\delta_1$ is interval of observation between these processes. Below are the illustrations of both double controls' intervals, and their impulse $u_\pm$ actions.

All illustrating intervals on the figure are expanded without their proper scaling.

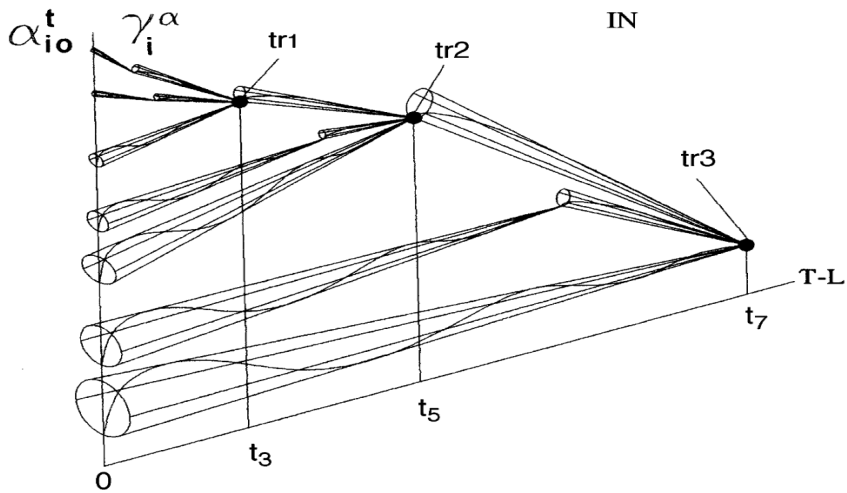

**Fig. 3.** The IN time-space information structure, represented by the hierarchy of the IN cones' spiral space-time dynamics with the triplet node's (tr1, tr2, tr3, ..), formed at the localities of the triple cones vertexes' intersections (knots), where $\{\alpha_{io}^t\}$ is a ranged string of the initial eigenvalues, cooperating around the $(t_1, t_2, t_3)$ locations; T-L is a time-space.

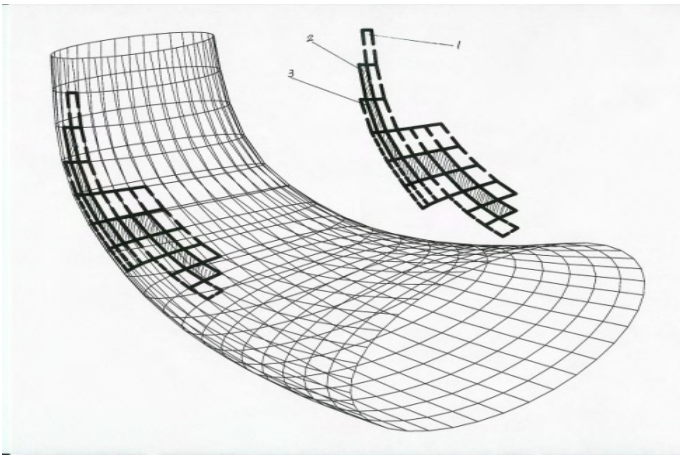

**Fig.4.** Illustration of the observer's self-forming cellular geometry by the cells of the DSS triplet's code, with a portion of the surface cells (1-2-3).

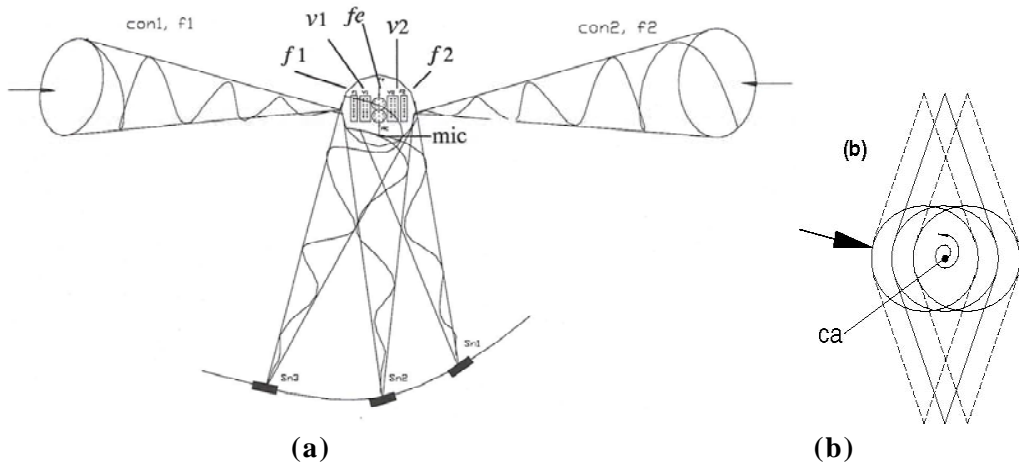

**Fig. 5**. An illustration of the competition and selection for the state's variations at the node locality, brought by: the external frequencies $f1, f2$, the controls $v1, v2$, the microlevel $mic$, and the external influences $fe$, with forming the triple node's spots $sn1, sn2, sn3$, generated by chaotic attractors on the cones basin; the node, cooperating the spots, is memorized.
b) Chaotic dynamics, generating an attractor (ca) at joining the cone's processes, which form the cones basin that emanates the spots.